\documentclass{webofc}
\usepackage[varg]{txfonts}   
\begin{document}
\title{Nucleosynthesis in Type Ia Supernovae, Classical Novae, and Type I X-Ray Bursts. A Primer on Stellar Explosions}
\author{\firstname{Jordi} \lastname{Jos\'e}\inst{1,2}\fnsep\thanks{\email{jordi.jose@upc.edu}} 
}

\institute{Departament de F\'\i sica, EEBE, Universitat Polit\`ecnica de Catalunya, c/ Eduard Maristany 16, E-08019 Barcelona, Spain
\and
           Institut d’Estudis Espacials de Catalunya, c/ Gran Capit\`a 2-4, Ed. Nexus-201, E-08034 Barcelona, Spain
          }

\abstract{%
Nuclear astrophysics aims at unraveling the cosmic origins of chemical elements and the physical processes powering stars.
It constitutes a truly multidisciplinary field, that integrates tools, advancements, and accomplishments from theoretical astrophysics, observational astronomy,
cosmochemistry, and theoretical and experimental atomic and nuclear physics.
For instance, the advent of high-energy astrophysics, facilitated by space-borne observatories, has ushered in a new era, offering a unique, panchromatic view of the universe
(i.e., allowing multifrequency observations of stellar events);
supercomputers are also playing a pivotal role, furnishing astrophysicists with computational capabilities essential for studying the intricate evolution of stars within a
multidimensional framework; cosmochemists, through examination of primitive meteorites, are uncovering tiny fragments of {\it stardust}, shedding light on the physical
processes operating in stars and on the mechanisms that govern condensation of stellar ejecta into solids; simultaneously, nuclear physicists managed to measure 
nuclear reactions at (or close to) stellar energies, using both stable and radioactive ion beam facilities.
This paper provides a multidisciplinary view on nucleosynthesis accompanying stellar explosions, with a specific focus on thermonuclear supernovae, classical novae, and type I X-ray bursts.
}
\maketitle
\section{Introduction}
\label{intro}
Stars are intricate nuclear furnaces, wielding a pivotal influence in the chemical enrichment of the universe.
The concept that stars are the crucibles where chemical elements are forged was formulated in the mid 1940s by Hoyle, building upon earlier work from
Bethe, Gamow, von Weizsäcker, and others in the 1920s and 1930s. This theory found empirical validation with the identification of technetium in the spectra of various S
stars---a type of stars that contain s-process elements in their spectra \cite{Mer52}. Remarkably, technetium lacks stable isotopes, its longest-lived one
boasting a half-life of approximately 4 million years. Hence, the discovery of this element served as irrefutable evidence that nucleosynthesis is ongoing (i.e., did not take
place exclusively at the early stages of the universe).
The subsequent discovery of other short-lived species further corroborated the key role played by stars in the synthesis of the elements.

This paper explores the pivotal role that stellar explosions play in the synthesis of cosmic elements.
Specifically, we will address how white dwarfs and neutron stars can be rejuvenated from a {\it posthumous} state through mass-transfer episodes from a stellar companion, 
giving rise to some of the most captivating stellar phenomena in the cosmos: type Ia supernovae, classical novae, and X-ray bursts.

\section{Type Ia supernovae}
\label{sec-1}
The prevailing taxonomy of supernovae relies heavily on optical spectroscopic measurements near maximum brightness, as discussed in refs. \cite{WH90,Fil97,Bra98}.
Broadly categorized into two main groups, supernovae are distinguished by the absence (type I) or presence (type II) of hydrogen (H) in their spectra.
Notably, the absence of H puts constraints on the maximum amount of this element in the expanding
atmosphere---roughly ${\rm M_H} \leq$ 0.03 - 0.1 M$_\odot$. Some SNIa (e.g., SN 2002ic), however, have revealed the presence of H emission lines some months
after achieving maximum brightness, whose origin has been traced to interactions between the supernova ejecta and H-rich circumstellar material.
SNIa also exhibit a distinctive absorption feature near 6150 \AA, due to blueshifted Si II, absent in subclasses Ib and Ic (which, in turn, showcase prominent
oxygen and sodium absorption lines). 
While SNIa are observed in all galaxy types, types Ib/c and II are only found in spiral and irregular galaxies, hinting at the likely association
of SNIa with old progenitors, while other classes must involve younger (and massive) stars. The energy released in a SNIa, approximately ${\rm E_{kin}} \sim 10^{51}$ erg, with expansion
velocities between 5000 and 10000 km s$^{-1}$, far exceeds the energy integrated over the light curve, ${\rm E_{rad}} \sim 10^{49}$ erg, revealing that what is observed
in a SNIa is basically the fallout from a thermonuclear explosion.
Early work by Hoyle and Fowler \cite{HF60} linked SNIa to the ignition of degenerate CO-rich white dwarfs.

Photometrically, SNIa showcase a rapid luminosity increase in around 20 days, peaking at ${\rm L_{peak}} \sim 10^{10}$ L$_\odot$. This is followed by a steep decline (by a factor
of $\sim 20$) in approximately 30 days, and later by a second, smoother decline over around 70 days \cite{Kir90,HW90,LP92}. Peak luminosities depend on the total synthesized $^{56}$Ni
(between 0.1 - 1 M$_\odot$), with late-stage light curve powered by the decay chain $^{56}$Ni $\rightarrow$ $^{56}$Co $\rightarrow$ $^{56}$Fe.
70\% of all observed SNIa share similar spectral features, peak luminosities, light curve shapes and characteristic timescales, favoring a dominant stellar progenitor and explosion
mechanism (most likely, the disruptive explosion of a 1.4 M$_\odot$ CO-rich white dwarf). However,  the increasing number of peculiar supernovae discovered in recent years
has spurred interest in alternative explosion mechanisms.
Differences in light curves have been attributed to varying amounts of synthesized $^{56}$Ni. In this context, two main scenarios have been proposed to explain the origin
of {\it normal} SNIa: a {\it single-degenerate channel}, 
involving transfer of H- or He-rich matter from a low-mass star onto a CO white dwarf, and a {\it double-degenerate channel}, where two CO white dwarfs merge as a result
of energy and angular momentum losses driven by the emission of gravitational waves. Both scenarios present challenges and advantages: the former faces 
difficulties in the white dwarf's evolutionary pathway to grow in mass and reach the Chandrasekhar limit, while the latter contends with the scarcity of likely candidates for a 
Chandrasekhar-mass explosion.

Rough estimates suggest that once per century in our Galaxy, and perhaps 10 per second in the observable universe, a white dwarf gets disrupted in a supernova explosion. Comparative analyses of
chemical abundances from SNIa spectra and theoretical predictions have helped to unveil the physical mechanisms behind these explosive events.
In this context, efforts have focused on understanding the nature of the burning front that propagates through (and incinerates) the star.
Pioneering numerical simulations of C-ignition \cite{Arn68,Arn69,Arn71} revealed that a supersonic (detonation) front would result in unrealistically high abundances of
iron-peak nuclei, and simultaneously, a low production of intermediate-mass elements (i.e., silicon, sulfur, and calcium), at odds with the abundance patterns inferred from the spectra.
Nomoto et al. \cite{Nom76,Nom84,Thi86} suggested that at the typical densities at which C is expected to ignite near the center of a white dwarf (approximately $\sim 10^9$ g cm$^{-3}$),
a subsonic (deflagration) burning front, catalyzed by the thermal conduction of the degenerate electron gas, is more likely to happen.  These C-deflagration models
showed a better alignment with the observed chemical abundances, successfully reducing the overproduction of iron-peak elements while enhancing the amounts of intermediate-mass species
like $^{40}$Ca, $^{36}$Ar, and $^{32}$S.  However, a notable drawback of these models is the severe overproduction of a number of n-rich isotopes, including $^{54}$Cr and $^{50}$Ti, 
in comparison to solar abundances.
In response to these shortcomings, new-generation SNIa models emerged in the 1990s, with delayed detonation models standing out as the most successful to date.
In these models, an early deflagration front propagates and preexpands the star, subsequently transitioning into a detonation front \cite{Kho01,Nie99}. The primary source of uncertainty
in these models lies in the physical mechanism driving the deflagration-to-detonation transition.

\subsection{Nucleosynthesis in type Ia supernovae}
Nucleosynthesis in SNIa depends critically on the peak temperature attained and on the specific density at the onset of the explosion.
The observed chemical abundance pattern is expected to reflect the interplay of five distinct burning regimes within the star, including normal and $\alpha$-rich freeze-out
from nuclear statistical equilibrium at the inner regions of the star, and incomplete Si-, O-, and C/Ne-burning in the outer layers \cite{Thi86,Woo86}.

Extensive studies of the sensitivity of predicted SNIa nucleosynthesis to variations in both thermonuclear
reaction and weak interaction rates have been performed in recent years, considering both C-deflagration and delayed-detonation models \cite{Par13a}.
These studies revealed the key role played by nuclear uncertainties affecting
a number of reactions, most notably  $^{12}$C($\alpha, \gamma)$, $^{12}$C+$^{12}$C, $^{20}$Ne($\alpha$, p), $^{20}$Ne($\alpha,\gamma$), and $^{30}$Si(p, $\gamma$).
Additionally, variations in the stellar $^{28}$Si($\beta^+)^{28}$Al, $^{32}$S($\beta^+)^{32}$P,
and $^{36}$Ar($\beta^+)^{36}$Cl rates were identified as the main weak interactions significantly affecting the final yields in any of the models considered.

\section{Classical novae}
\label{sec-2}
Classical novae (CN) represent another class of stellar explosions occurring in binary systems, featuring a white dwarf (typically CO- or ONe-rich) and a low-mass
main sequence or more evolved companion. These events exhibit a rapid rise in optical brightness within 1 - 2 days, peaking at luminosities ranging from $10^4$ to $10^5$ L$_\odot$.
The binary systems leading to nova outbursts are characterized by short orbital periods (i.e., less than 15 hours), enabling mass-transfer episodes due to Roche Lobe overflow
of the secondary star. Since the transferred material carries angular momentum, it forms an accretion disk around the white dwarf. Ultimately, a portion of this material spirals in,
accumulating on the white dwarf's surface, bulding up an envelope under mildly degenerate conditions until a thermonuclear runaway occurs \cite{JS08,Jos16}.

Nova explosions are relatively common in the universe and constitute the second most frequent type of stellar thermonuclear explosions in our Galaxy after type I X-ray bursts
(see Sect. \ref{sec-3}), estimated to occur at a rate of 50 yr$^{-1}$ \cite{Sha17}. However, their detection from Earth is limited by interstellar extinction from dust,
resulting in the discovery of only a fraction, around 5 - 10 per year.

In contrast to the complete disruption of the star expected in type Ia supernovae, neither the white dwarf nor the binary system is destroyed by a nova explosion.
As a result, CN events are expected to recur, typically after $10^4$ - $10^5$ years. Recurrent novae, by definition novae  observed in outburst more than once, display a shorter recurrence
time of only 1 - 100 years, implying very massive white dwarfs close to the Chandrasekhar-mass limit and high mass-accretion rates. Whether the range of recurrence times follows
a continuous sequence, spanning from the shorter values characterizing recurrent novae 
to the longer values predicted for classical novae (i.e., 1 - $10^5$ yr), remains a topic of debate.
Another significant distinction between CN and SNIa lies in the velocity of the ejecta (exceeding $10^4$ km s$^{-1}$ in SNIa compared to several $10^3$ km s$^{-1}$ in CN)
as well as in the amount of mass ejected (the entire star, approximately 1.4 M$_\odot$, in a thermonuclear supernova versus 10$^{-7}$ - 10$^{-4}$ M$_\odot$ for a nova).

Mass-accretion leads to material accumulation on the white dwarf's surface, causing compressional heating and triggering nuclear reactions. The huge energy
released cannot be solely radiated away, leading to the onset of convection once superadiabatic gradients establish in the accreted envelope. Convection redistributes short-lived
$\beta^+$-unstable nuclei, such as $^{13}$N, $^{14,15}$O, and $^{17}$F, synthesized deep inside the envelope, to the outer, cooler regions. A portion of the energy released by the
$\beta^+$-decay of these species transforms into kinetic energy, propelling the ultimate expansion and ejection stages of the nova.
The strength of the resulting outburst depends critically on four parameters: the mass and initial luminosity (or temperature) of the white dwarf hosting
the explosion, and the metallicity and mass-transfer rate from its stellar companion. Notably, the runaway is halted by envelope expansion rather than by fuel consumption,
in sharp contrast to type I X-ray bursts (see Sect. \ref{sec-3}).

Numerical simulations have revealed that envelopes with solar metallicity can lead to explosions resembling "slow novae" \cite{SST78,Pri78}. However, to replicate the
overall observational characteristics of a "fast nova", only envelopes with enhanced CNO abundances (in the range Z$_{\rm CNO} \sim 0.2$ - $0.5$) are effective \cite{Sta72,Sta78b}.
The origin of such CNO enhancements, both required by models and spectroscopically inferred, has been a subject of controversy. 
Two possibilities have been considered: nuclear processing during the explosion or mixing at the core-envelope interface.
Peak temperatures attained during a nova explosion, constrained by the chemical abundance pattern in the ejecta, do not exceed $4 \times 10^8$ K.
Hence, it is improbable that the observed metallicity enhancements can be attributed to thermonuclear processing driven by CNO breakout.
Instead, mixing at the core-envelope interface emerges as a more plausible explanation. Various mixing mechanisms, including diffusion-induced mixing,
shear mixing at the disk-envelope interface, convective overshoot-induced flame propagation \cite{LT90}, or mixing by gravity wave breaking on the white dwarf
surface \cite{Ros01,Ale04}, have been proposed and explored, but none has proven entirely successful.
Significant progress has been made when relaxing the constraints imposed by spherically symmetric models.
Multidimensional simulations of mixing at the core-envelope interface during nova outbursts have proved that Kelvin-Helmholtz instabilities can naturally
lead to self-enrichment of the accreted envelope with core material, aligning with observations. Pioneering 3-D simulations \cite{Cas11b,Cas16,Jos20} have provided insights
into the nature of the highly fragmented, chemically enriched, and inhomogeneous nova shells observed in high resolution. These patterns, predicted by the Kolmogorov theory
of turbulence, are interpreted as relics of the hydrodynamic instabilities that develop during the initial ejection stage. While the inhomogeneous patterns inferred
from the ejecta have been traditionally considered as potentially arising from observational uncertainties,
they may actually represent a genuine signature of the turbulence generated during the thermonuclear runaway.

\subsection{Nucleosynthesis in classical nova outbursts}
From a nuclear physics perspective, the initial phases of the thermonuclear runaway characterizing a nova outburst are primarily driven by the proton-proton chains
and the cold CNO cycle (i.e., $^{12}$C(p, $\gamma$)$^{13}$N($\beta^+$)$^{13}$C(p, $\gamma$)$^{14}$N). As the temperature rises, the characteristic timescale for proton captures
onto $^{13}$N becomes shorter than the corresponding $\beta^+$-decay time. This favors a series of reactions within the hot CNO cycle, including $^{13}$N(p, $\gamma$)$^{14}$O,
as well as $^{14}$N(p, $\gamma$)$^{15}$O and $^{16}$O(p, $\gamma$)$^{17}$F.
The substantial production of $^{13}$N, $^{14,15}$O, and $^{17}$F during the outburst translates into significant quantities of their daughter nuclei $^{13}$C, $^{15}$N, and $^{17}$O
in the ejecta. These elements constitute the primary contribution of novae to Galactic abundances.

The main nuclear path during nova outbursts runs close to the valley of stability, and is driven by proton-capture reactions and $\beta^+$-decays.
Classical novae are unique stellar explosions, since their nuclear activity, limited to approximately a hundred relevant species (with mass number A $<$ 40)
connected through a few hundred nuclear reactions, as well as the  moderate temperatures achieved during the explosion ($10^7$ - $4 \times 10^8$ K),
allow reliance primarily on experimental information \cite{JHI06}.
Numerous studies have aimed to identify the critical reactions whose uncertainties have a significant impact on nova nucleosynthesis.
Recent re-evaluations have addressed many of these reactions, and only uncertainties associated with the reactions $^{18}$F(p, $\alpha$)$^{15}$O,
$^{25}$Al(p, $\gamma$)$^{26}$Si, and $^{30}$P(p, $\gamma$)$^{31}$S still have a strong impact on nova nucleosynthesis.

Several species synthesized during classical nova outbursts may emit potentially detectable $\gamma$-rays.
This includes $^{13}$N and $^{18}$F, which power prompt $\gamma$-ray emission at and below 511 keV. Additionally, longer-lived isotopes such as $^{7}$Be and $^{22}$Na
decay when the envelope becomes optically thin to $\gamma$-rays, producing line emission at 478 and 1275 keV, respectively.
Although $^{26}$Al is synthesized during nova outbursts, only its cumulative emission can be observed due to its slow decay.
However, the contribution of novae to the Galactic content of $^{26}$Al is expected to be relatively small (less than 20\%) \cite{JHC97}.

Comparisons between abundance patterns inferred from observations and theoretical predictions generally show good agreement, with a predicted nucleosynthetic endpoint
around calcium. However, spectroscopically inferred abundance patterns yield only atomic values, limiting the direct comparison with theoretical predictions.
Laboratory analyses of presolar meteoritic grains provide better perspectives. Infrared and ultraviolet observations often reveal dust formation episodes in the nova ejecta
\cite{Geh98}. Since the pioneering studies of dust formation in novae by Clayton and Hoyle \cite{CH76}, efforts to identify presolar nova candidate grains have focused on 
low $^{20}$Ne/$^{22}$Ne ratios, as noble gases like
Ne do not condense into grains. The presence of $^{22}$Ne is attributed to in-situ $^{22}$Na decay, a signature of a classical nova explosion.
Notable progress has been made in identifying presolar nova candidate grains, including SiC and graphite grains with isotopic signatures consistent with
nova model predictions \cite{Ama01, Ili18, Hae19}.

\section{Type I X-ray bursts}
\label{sec-3}
X-ray bursts (XRBs) have been discovered more recently compared to novae and supernovae, as their major energy output is in X-rays, requiring space observatories for detection
(they are optically faint objects). Over 100 Galactic low-mass X-ray binaries exhibiting such bursting behavior have been found 
since XRBs were first discovered \cite{Bab75,Gri76,Bel76}. 
These events, resembling CNe, are hosted by neutron stars (with a mass up to $\sim 2 - 3$ M$_\odot$, and a very small diameter, 20 to 30 km), resulting from type II supernova
explosions of stars more massive than 10 M$_\odot$ (even though neutron stars can also form in other astrophysical scenarios, as in the accretion-induced collapse of a white dwarf).
The companion is often a main-sequence star or a red giant. XRBs are also recurrent events with much shorter recurrence periods than novae, ranging typically from hours to days.

XRBs are characterized by a rapid increase in brightness, reaching L${\rm peak} \sim 10^4 - 10^5$ L$_\odot$ after a very fast rise (in $\sim$ 1 to 10 seconds).
Their light curves are described by the ratio of persistent over burst luminosities ($\alpha \sim 100$).
The overall energy output in a typical XRB is $\sim 10^{39}$ erg, released over 10 to 100 seconds.

The potential contribution of XRBs to galactic abundances remains contentious,
because of the extremely large escape velocities from a neutron star surface. Indeed, the energy required to escape from the strong gravitational field of a neutron star
of mass M and radius R is G M m$_p$ /R $\sim$ 200 MeV/nucleon, whereas the nuclear energy released from thermonuclear
fusion of solar-like matter into Fe-group elements is only about 5 MeV/nucleon. 
Despite the improbable ejection of matter during the explosion, radiation-driven winds may eject a minute fraction (less than 1\%) of the envelope, contributing to the 
Galactic metal enrichment in elements around mass 60 
\cite{Herre20,Herre23}. 
This needs to be investigated in detail, given the conflicting results from attempts to identify chemical species through 
gravitationally-redshifted absorption lines, initially on the basis of high-resolution spectra of 28 XRBs detected from the source EXO 0748-676 after 335 ks of observations with the XMM-
Newton X-ray satellite \cite{Cot02}. This preliminary study identified lines of Fe XXVI (during the early phase of the bursts), Fe
XXV, and perhaps O VIII (at later stages). But no evidence for such spectral features was found neither during the
analysis of 16 bursts observed from GS 1826-24 \cite{Kong07}, nor from another series of bursting episodes detected from
the original source after 600 ks of observations \cite{Cot08}. It is, however, worth noting that another study identified
strong absorption edges in two XRBs exhibiting strong photospheric expansion. The spectral features were attributed
to Fe-peak elements with abundances about 100 times solar, which may suggest the presence of heavy-element ashes
in the ejected wind \cite{iW10}. This issue clearly deserves further theoretical and observational work.

\subsection{Nucleosynthesis in type I X-ray bursts}
During accretion onto neutron stars, the strong surface gravity leads to high temperatures and densities in the accreted envelopes, about an order of magnitude
larger than in a typical nova outburst. 
The main nuclear reaction flow in XRBs is dominated by the rp-process (rapid proton-captures and $\beta^+$-decays), together with the 3$\alpha$ reaction, and the $\alpha$p-process
 (a sequence of ($\alpha$,p) and (p,$\gamma$) reactions). 
The flow proceeds far away from the valley of stability, even merging with the proton drip-line beyond mass A = 38 \cite{Sch99,Sch01,Woo04,Fis08,Jos10}.

Most of the reaction rates used in XRB nucleosynthesis studies have been derived from statistical models, thus carrying notable uncertainties. Various studies 
have been made to assess the impact of such nuclear uncertainties by different research groups. For example, Parikh et al. \cite{Par08,Par09} used two distinct approaches.
In a first stage, they scrutinized the effect of individual reaction-rate variations through post-processing calculations using several temperature and density versus time profiles. 
To achieve this, an extensive nuclear network comprising 606 isotopes (ranging from H to $^{113}$Xe) was adopted.
Only a handful of reactions, out of the 3551 nuclear processes
considered, resulted in a significant impact on the final yields, when their nominal rates were varied by a factor of 10,
up and down. This included mostly proton-capture reactions, such as
 $^{65}$As(p, $\gamma$)$^{66}$Se, $^{61}$Ga(p, $\gamma$)$^{62}$Ge, $^{96}$Ag(p, $\gamma$)$^{97}$Cd, $^{59}$Cu(p, $\gamma$)$^{60}$Zn, $^{86}$Mo(p, $\gamma$)$^{87}$Tc, $^{92}$Ru(p, $\gamma$)$^{93}$Rh, or $^{102,103}$In(p, $\gamma$)$^{103,104}$Sn, as well as a few $\alpha$-capture reactions like $^{12}$C($\alpha$, $\gamma$)$^{16}$O, $^{30}$S($\alpha$, p)$^{33}$Cl, or $^{56}$Ni($\alpha$, p)$^{59}$Cu.
Additionally, the study identified certain reactions that significantly impact energy production when varied within a factor of 10. This underscores the limitations of postprocessing techniques. 
A self-consistent analysis would require computationally intensive hydrodynamic simulations capable of self-adjusting both the temperature and density of the stellar envelope.
Similar results were obtained in a subsequent study based on a Monte Carlo approach, where all reaction rates were simultaneously varied by random factors. Remarkably, all reactions deemed significant in the Monte Carlo study had been previously identified in the individual reaction-rate variation analysis. 

\section{ACKNOWLEDGMENTS}
This work has been partially supported by the Spanish MINECO grant PID2020-117252GB-I00, by the E.U. FEDER funds, and
by the AGAUR/Generalitat de Catalunya grant SGR-386/2021. This article benefited from discussions within the EU H2020 project No. 101008324
“ChETEC-INFRA”.
\bibliographystyle{aipnum-cp}%
\bibliography{jjose18v4}%

\end{document}